\documentstyle[prl,aps,twocolumn,psfig]{revtex}

\begin{document}
\twocolumn[\hsize\textwidth\columnwidth\hsize\csname
@twocolumnfalse\endcsname 


\tighten
\draft
\title{Cosmological Constraints on Late-time Entropy Production}
\author{M. Kawasaki and  K. Kohri}
\address{Research Center for the Early
Universe, Faculty of Science, The University of Tokyo, Tokyo 113-0033,
Japan}
\author{Naoshi Sugiyama}
\address{Department of Physics, Kyoto University, Kyoto 606-8502, Japan}

\maketitle
\begin{abstract}
    
    We investigate cosmological effects concerning the late-time
    entropy production due to the decay of non-relativistic massive
    particles. The thermalization process of neutrinos after the
    entropy production is properly solved by using the
    Boltzmann equation.  If a large entropy
    production takes place at late time t$\simeq$ 1 sec, it is found
    that a large fraction of neutrinos cannot be thermalized.  This
    fact loosens the tight constraint on the reheating temperature
    $T_{R}$ from the big bang nucleosynthesis and $T_{R}$ could be as
    low as 0.5 MeV. The influence on the large scale structure
    formation and cosmic microwave background anisotropies is also
    discussed.

\end{abstract}

\pacs{98.80.Cq, 98.70.Vc 
\hspace{1cm} RESCEU 7/99, KUNS-1546, astro-ph/9811437}

]


It is usually believed that thermal radiation dominates the energy
density of the early universe after the reheating process of the
primordial inflation. At least, the universe is expected to be
radiation-dominated before the big bang nucleosynthesis (BBN) epoch
($t \simeq 1$ sec) otherwise the abundances of the synthesized light
elements ($^4$He, $^3$He, D and $^7$Li) do not agree with
observations~\cite{Kolb-Turner}. However, it is uncertain that the
universe is radiation dominated before BBN epoch.  In fact, particle
physics models beyond the standard one predict a number of new massive
particles some of which have long lifetimes and may influence the
standard BBN scenario.  Since the energy density of such massive
non-relativistic particles decreases more slowly than that of the
radiation as the universe expands, the universe becomes
matter-dominated by those particles until they decay.  When they decay
into ordinary particles, the large entropy is produced and universe
becomes radiation-dominated again. We call this process ``late-time
entropy production''.

One can find some interesting candidates for the late-time entropy
production in models based on supersymmetry (SUSY). In local SUSY (i.e.
supergravity ) theories~\cite{Nilles} there exist gravitino and Polonyi
field~\cite{Polonyi} which have masses of $\sim {\cal O}(100{\rm
  GeV}-10{\rm TeV})$. Since gravitino and Polonyi field interact with
other particles only through gravity, they have long lifetimes. For
example, the Polonyi field with mass $\sim 10$~TeV quickly dominates the
energy density of the universe because this field cannot be diluted by
usual inflation and decays at BBN epoch. It is also known that
superstring theories have many light fields such as dilaton and moduli
which have similar properties as the Polonyi field.

When one considers the late-time entropy production, reheating
temperature $T_{R}$ is usually used as a parameter to characterize it.
The reheating temperature $T_{R}$ is determined from $\Gamma = 3
H(T_{R})$ where $\Gamma$ is the decay rate and $H(T_{R})$ is the
Hubble parameter at the decay epoch. Since the Hubble parameter is
expressed as $H = \sqrt{g_{*}\pi^2/90}T_{R}^2/M_G$, where $g_{*}$ is
the number of massless degrees of freedom (= 43/4) and $M_G$
is the reduced Planck mass ($= 2.4\times 10^{18}$GeV), the reheating
temperature is estimated as $ T_{R} = 0.554 \sqrt{\Gamma M_G}$.


As mentioned above, the stringent constraint on the late-time entropy
production or reheating temperature comes from the consideration of
BBN. The long-lived massive particles which are responsible for the
late-time entropy production should decay early enough to make the
universe to be dominated by thermal radiation before the BBN epoch. To
establish the thermal equilibrium, the decay products should be
quickly thermalized through scatterings, annihilations, pair creations
and further decays. Almost all standard particles except neutrinos are
thermalized very soon when they are produced in the decay and
subsequent thermalization processes.  Neutrinos can be thermalized
only through weak interaction which usually decouples at a few MeV.
Thus, the thermalization of neutrinos is most important to obtain
constraints on the reheating temperature. However, the thermalization
of neutrinos has not been well studied and people have used various
constraints on the reheating temperature between 1MeV and 10MeV.
Therefore, in this letter, we will obtain the constraint on the
reheating temperature by using the neutrino spectrum obtained from
numerical integration of a set of Boltzmann equations together with
full BBN network calculations.

Another interesting constraint may come from observations of
anisotropies of the cosmic microwave background radiations (CMB). It is
known that the CMB anisotropies are
very sensitive to the time of matter-radiation equality 
(see e.g., \cite{Lopez}) . When the
reheating temperature is so low that sufficient neutrinos cannot be
thermally produced, the radiation ($=$ photons $+$ neutrinos) density
becomes less than that in the standard case, which may give
distinguishable signals in the CMB anisotropies. In this letter, we
use the effective number of neutrino species $N_{\nu}^{\rm eff}$ as a
parameter which represents the the energy density of neutrinos 
defined by
$    N_{\nu}^{\rm eff} 
    \equiv {\sum_i\rho_{\nu_i}}/{\rho_{\rm std}}, 
$
where $i=\nu_{e}, \nu_{\mu}, \nu_{\tau}$, and $\rho_{\rm std}$ is the
neutrino energy density in the standard case (i.e. no late-time
entropy production). 

First, let us discuss the neutrino spectrum. When a massive particle
$\phi$ which is responsible for the late-time entropy production
decays, all emitted particles except neutrinos are quickly thermalized
and make a thermal plasma with temperature $\sim T_{R}$. If the
reheating temperature is high enough ($T_{R} \gg 10$ MeV), there is no
question about the neutrino thermalization. For relatively low
reheating temperature ($T_{R} \lesssim 10$MeV), however, neutrinos are
slowly thermalized and may not be in time for the beginning of BBN. We
assume that the decay branching ratio into neutrinos is very small and
that neutrinos are produced only through annihilations of electrons
and positrons, i.e. $e^{+} + e^{-} \rightarrow \nu_{i} + \bar{\nu}_{i}
(i= e, \mu, \tau)$. The evolution of the distribution function $f_{i}$
of the neutrino $\nu_{i}$ is described by the Boltzmann equation:
\begin{equation}
    \label{eq:Boltzmann}
    \frac{\partial f_{i}(p_{i},t)}{\partial t} 
    - H p_{i}\frac{\partial f_{i}(p_{i},t)}{\partial p_{i}}
    =  C^{\rm a}_{i} + C^{ \rm s}_{i},
\end{equation}
where $p_{i}$ is the momentum of $\nu_{i}$ and $C^{\rm a}_{i} (C^{ \rm
s}_{i})$ is the collision term for annihilation (scattering)
processes.  Here we consider the following processes: $ \nu_{i} +
\nu_{i} \leftrightarrow e^{+} + e^{-}$ and $ \nu_{i} + e^{\pm}
\leftrightarrow \nu_{i} + e^{\pm}$. We do not include the neutrino
self interactions which may not change the result in the present
paper, since the neutrino number densities are much smaller than the
electron number density at low reheating temperature.

Here we have treated neutrinos as Majorana ones (i.e., $\nu =
\bar{\nu}$).  Note that our results are the same for Dirac neutrinos.
The collision terms are quite complicated and expressed by nine
dimensional integrations over momentum space. However, if we neglect
electron mass and assume that electrons obey the Boltzmann
distribution $e^{-p/T}$, the collision terms are simplified to one
dimensional integration forms. Because the weak interaction rate is
small at T $\lesssim$ 0.5 MeV, neglecting the electron mass changes
the result little. Then, $C^{\rm a}_{i}$ is given by~\cite{Kawasaki}
\begin{eqnarray}
    \label{eq:C-ann}
    C^{\rm a}_{i} =   
   - \int \frac{dp'_i}{\pi^2} p'^2_i
    (\sigma v)_i (f_i(p_i) f_i(p'_i)- f_{eq}(p_i)f_{eq}(p'_i)), 
\end{eqnarray}
where $f_{eq} (= 1/(e^{p_i/T} +1))$ is the equilibrium distribution and
$(\sigma v)_i$ is the differential cross sections given by
$    (\sigma v)_{e}  =  (4G^2_F/(9\pi))
    ((C_V + 1)^2 + (C_A + 1)^2)pp',$ and 
$    (\sigma v)_{\mu,\tau}  =  (4G^2_F/(9\pi))
    (C_V^2 + C_A^2)pp'$.
Here $G_F$ is the Fermi coupling constant, and $C_A = -1/2, C_V = -1/2 +
2 \sin^2\theta_W$ ($\theta_W$: Weinberg angle). 

As for scattering processes, $C^{ \rm s}_{i}$ is expressed as
\begin{eqnarray}
    \lefteqn {C^{ \rm s}_{i} = \frac{2G_F^2}{\pi^3}(C_V^2+C_A^2)\times}
    \nonumber \\
    & &\left[ - \frac{f_i}{p_i^2}
      \left(   \int_0^{p_i} dp'_i
        (1-f_i(p'_i))F_1 + \int^{\infty}_{p_i}dp'_i
        (1-f_i(p'_i))F_2\right)\right.
    \nonumber \\
    \label{eq:C-scat}
     & & + \left. \frac{1-f_i(p_i)}{p_i^2}
      \left(\int_0^{p_i}dp'_i f_i(p'_i)B_1
     + \int^{\infty}_{p_i}dp'_i f_i(p'_i)B_2\right)\right].
\end{eqnarray}
Here $(C_V^2+C_A^2)$ is replaced by $((C_V + 1)^2 + (C_A +
1)^2)$ for $i=e$, and  the functions $F_1, F_2, B_1, B_2$ are given by
\begin{eqnarray}
    F_1(p, p') & = & D(p, p') + E(p, p')e^{-p'/T}, 
    \nonumber\\
    F_2(p, p') & = & D(p', p)e^{(p-p')/T} + E(p, p')e^{-p'/T}, 
    \nonumber\\
    B_1(p,p')  & = & F_2(p',p), ~~B_2(p,p')   =  F_1(p',p), 
\end{eqnarray}
where 
\begin{eqnarray}
    D(p, p') & = & 2T^4(p^2 + p'^2 + 2T(p-p')+4T^2), 
    \nonumber\\ 
    E(p, p') & = & - T^2[p^2p'^2 + 2pp'(p+p')T 
    \nonumber\\ 
    & & + 2(p+p')^2T^2 + 4(p+p')T^3 + 8T^4].
\end{eqnarray}

Together with the above Boltzmann equations, we also solve the
evolution of the densities of the massive particle $\phi$, radiations
$\rho_r (= \rho_{\gamma}+\rho_{e^{\pm}})$ and the scale factor $a$:
\begin{eqnarray}
    \frac{d\rho_{\phi}}{dt} & = & -\Gamma\rho_{\phi} -3H\rho_{\phi},\\
    \frac{d\rho_r}{dt}&  = & \Gamma \rho_{\phi} 
    - \frac{d\rho_{\nu}}{dt} - 4H(\rho_r + \rho_\nu) , \\
    H & = & \frac{d\ln a}{dt} = \frac{1}{\sqrt{3}M_G}
    (\rho_{\phi} + \rho_{r} + \rho_{\nu})^{1/2},
\end{eqnarray}
where the neutrino density is given by $\rho_{\nu} = \sum_i 1/\pi^2
\int dp_i p_i p_i^3 f_i(p_i)$.

\begin{figure}[htbp]
    \centerline{\psfig{figure=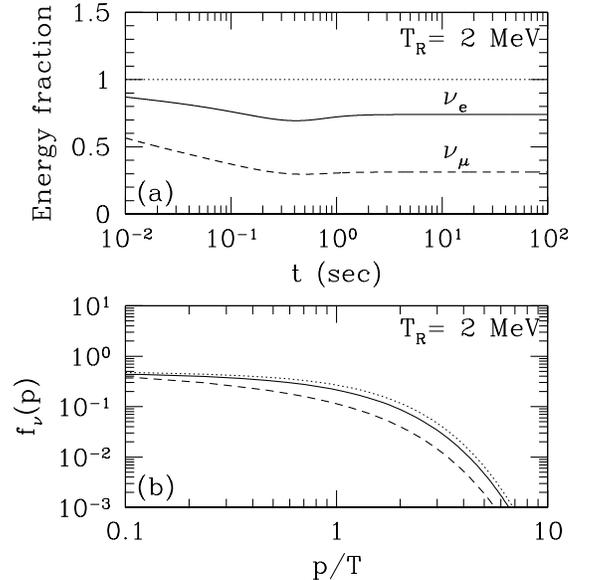,width=8.0cm}}
      \caption{
      (a) Evolution of the the energy density of $\nu_{e}$ (solid curve) and
      $\nu_{\mu}$ (dashed curve) for $T_{R}=2$~MeV. (b) Distribution of
      $\nu_{e} $ (solid curve) and $\nu_{\mu}$ (dashed curve) for
      $T_{R}=2$~MeV. The dotted curve is the thermal equilibrium Fermi-Dirac 
      distribution.}      \label{fig:rho-nu}
\end{figure}

In Fig.~\ref{fig:rho-nu}~(a) we show the evolutions of $\rho_{\nu_e}$ and
$\rho_{\nu_{\mu}}=\rho_{\nu_{\tau}}$ for $T_{R} = 2$~MeV.
Since the electron neutrinos interact with electrons or positrons
through both charged and neutral currents, they are more effectively
produced from the thermal plasma than the other neutrinos which have
only neutral current interactions.  
The final distribution functions $f_e$ and $f_{\mu}=f_{\tau}$ are
shown in Fig.~\ref{fig:rho-nu}~(b), from 
which one can see that the occupation numbers are close to
equilibrium values at low momentum but they deviate significantly at
higher momentum.

\begin{figure}[htbp]
    \centerline{\psfig{figure=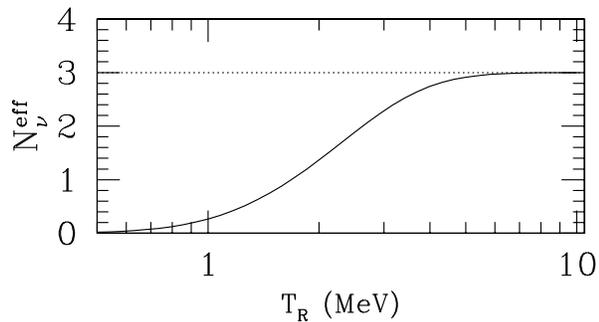,width=8.5cm}}
    \vspace{-4.0cm}
      \caption{
      $N_{\nu}^{\rm eff}$ as a function of $T_{R}$.}
      \label{fig:tr_nnu}
\end{figure}

In Fig.~\ref{fig:tr_nnu} we can see the change of $N_{\nu}^{\rm eff}$ as a
function of $T_{R}$. If $T_{R} \gtrsim 7$ MeV, $N_{\nu}^{\rm eff}$ is
almost equal to three
and neutrinos are thermalized very well. On the other hand, if $T_{R}
\lesssim 7$ MeV, $N_{\nu}^{\rm eff}$ becomes smaller than three. 

The deficit of the neutrino distribution influences the produced light
element abundances. In particular, the abundance of the primordial $^4$He
is drastically changed.  At the beginning of BBN (T $\sim$ 1 MeV - 0.1
MeV) the competition between the Hubble expansion rate and the weak
interaction rates determines the freeze-out value of neutron to proton
ratio. After the freeze-out time, neutrons can change into protons only
through the free decay with the life time $\tau_n $. Since the left
neutrons are almost included into $^4$He, the primordial $^4$He is
sensitive to the freeze-out value of neutron to proton ratio.

If the neutrino energy density gets smaller than that of the standard
BBN (SBBN), Hubble parameter is also decreased. Then the $\beta$
equilibrium between neutrons and protons continues for longer time,
and less neutrons are left. Thus, the predicted $^4$He is less than
the prediction of SBBN.  This effect is approximately estimated by:
$\Delta Y \simeq - 0.1 (- \Delta \rho/\rho)$, where $Y$ is the mass
fraction of $^4$He and $\rho$ is the total energy density of the
universe.

In addition if the electron neutrino is not thermalized sufficiently
and does not have the perfect Fermi-Dirac distribution, there are two
interesting effects by which more $^4$He are produced. The weak
reaction rates are computed by the neutriono distributions which are
obtained by solving Boltzmann equations. For example, a reaction rate
at which neutron is changed into proton ($n + \nu_e \to p + e^-$) is
represented by
\begin{eqnarray}
    \label{beta_reac}
    \Gamma_{n \nu_e \to p e^-} & = & 
    K \int_0^{\infty}dp_{\nu_e}\left[\sqrt{(Q+p_{\nu_e})^2-m_e^2}
    (p_{\nu_e}+Q)  
    p_{\nu_e}^2\right. \nonumber \\
    & & \left.\times
      \left(1-\frac{1}{e^{(p_{\nu_e}+Q)/T_{\gamma}}+1}\right) 
    f_{\nu_e}(p_{\nu_e})\right],
\end{eqnarray}
where $Q = m_n - m_p = 1.29$ MeV and $K$ is a normalization factor
which is determined by the neutron life time $\tau_n $ as $ K \simeq
(1.636 \tau_n)^{-1}$.  In this equation we can see that if the (anti)
electron neutrino distribution functions decrease, the weak
interaction rates also decrease.\footnote{The weak interaction rates
with (anti) electron neutrino in the final state slightly increase because the
Pauli blocking factor (1-$f_{\nu_e}$) increases. However, both of the
total weak interaction rates $\Gamma_{n\to p}$ and $\Gamma_{p\to n}$ decrease.}
 First, when the weak interaction rate
$\Gamma_{n\leftrightarrow p}$ decreases, the Hubble expansion becomes
more rapid than that of the interaction rate earlier. Then the
freeze-out value of neutron to proton ratio becomes larger than in
SBBN and the predicted $^4$He abundance becomes larger: $\Delta Y
\simeq + 0.15 (- \Delta\Gamma_{n\leftrightarrow
p}/\Gamma_{n\leftrightarrow p})$.  Second when the interaction rates
$\Gamma_{n \to p}$ at which neutrons are changed into protons become
smaller, less neutrons can turn into protons after the freeze-out
time. Then the produced $^4$He also becomes larger: $\Delta Y \simeq +
0.2 (- \Delta\Gamma_{n\to p}/\Gamma_{n\to p})$.
 
As for the observational abundances, we adopt the following values.
The primordial $^4$He mass fraction $Y$ is observed in the low
metallicity extragalactic HII regions. Now we have two observational
values, low $^4$He and high $^4$He, which are reported by different
groups.  We take ``Low $^4$He'' from Olive, Skillman and Steigman
(1997) \cite{OliSkiSte}, $Y^{obs}=0.234 \pm (0.002)_{stat} \pm
(0.005)_{syst}$.  Recently Izotov et al. \cite{Izo} claimed that the
effect of the HeI stellar absorption which are not considered well in
\cite{OliSkiSte} is very important. We adopt their value as ``High
$^4$He'', $Y^{obs}=0.244 \pm (0.002)_{stat} \pm (0.005)_{syst}$.

The deuterium D/H is measured in the high redshift QSO
absorption systems. Here we adopt the most reliable data
D/H $= (3.39 \pm 0.25) \times 10^{-5}$ \cite{BurTyt}.

The $^7$Li/H is observed in the Pop II old halo stars. We take the
recent measurements \cite{BonMol} and adopt the additional larger
systematic error, for fear there are underestimates in the stellar
depletion and the production by the cosmic ray spallation. Then we
obtain: Log$_{10}$($^7$Li/H) $=-9.76 \pm (0.012)_{stat} \pm
(0.05)_{syst} \pm (0.3)_{add}$.

\begin{figure}[htbp]
    \centerline{\psfig{figure=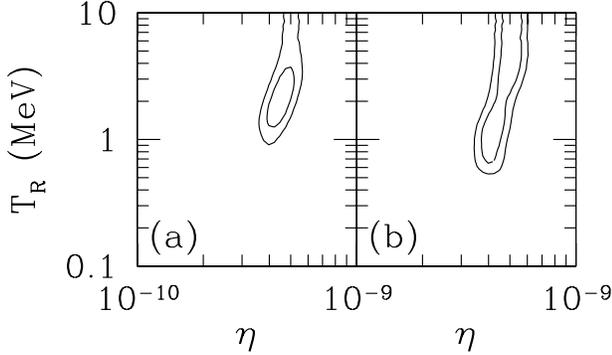,width=8.5cm}}

    \vspace{-3.cm}

      \caption{
       Contours of the confidence levels in ($\eta,T_R$) plane for 
      (a) observational value of Low $^4$He and (b) High $^4$He. The
      inner (outer) curve is  68$\%$ (95$\%$) C.L..
      }
      \label{fig:eta_tr}
\end{figure}

In order to discuss how the theoretical predictions with the late-time
entropy production agree with the above observational constraints, we
perform Monte Carlo calculation and the maximum likelihood analysis [
for details, e.g. see \cite{HKKM}].  In Fig.~\ref{fig:eta_tr} we plot
the contours of the confidence level in the $\eta$-T$_R$ plane for (a)
Low $^4$He and (b) High $^4$He. As we mentioned above, for T$_R
\gtrsim 7$ MeV, the theoretical prediction is the same as SBBN. On the
other hand as T$_R$ decreases, $Y$ gradually becomes smaller because
the effective number of neutrino species $N_{\nu}^{\rm eff}$
decreases. On the other hand, for T$_R \lesssim 2$ MeV the effect that
the weak interaction rates are weakened due to the lack of the
neutrino distributions begins to be important and $Y$ begins to
increase as T$_R$ decreases. For T$_R \lesssim 1$ MeV, the weak
interaction rates are still more weakened and $Y$ steeply increases as
$T_R$ decreases because it is too late to produce enough electrons and
positrons whose mass is about $m_e$ = 0.511 MeV.  From
Fig.~\ref{fig:eta_tr} we can conclude that $T_R \lesssim 0.5$~MeV is
excluded at 95 $\%$ C.L.  Or we could rather say that $T_R$ can be as low as 
$0.5\rm MeV$. Accordingly $N_{\nu}^{\rm eff}$ can be as small as $0.1$.

\vspace{-0.5cm}

The late time entropy production is also constrained from 
the formation of the large scale structure of the universe.
We can fit the large scale galaxy distribution if 
the ``shape parameter'' $\Gamma_{\rm s} \simeq 0.25\pm 0.05$~\cite{PD}.
In case of the standard cold dark matter (CDM) scenario,
$\Gamma_{\rm s} \equiv \Omega_0 h$ whose dependence is 
determined by the epoch of the matter-radiation equality.
Here $\Omega_0$ is the density parameter and $h$ is the non-dimensional 
Hubble constant normalized by $100\rm km/s/Mpc$.
Since $\Gamma_{\rm s} \sim 0.5$ for so-called standard CDM model, 
it is known that this model is in serious trouble and one can 
achieve the desired fit with low density models where 
$\Omega_0 \simeq 0.3$.  If the late time entropy production 
takes place, however,
$\Gamma_{\rm s}$ is modified as 
$\Gamma_{\rm s} = 1.68 \Omega_0 h/(1+0.23N_\nu^{\rm eff})$.
Therefore the constraint from the large scale galaxy distribution becomes 
much tighter with $N_\nu^{\rm eff} < 3$  (see Fig. 4).   
\begin{figure}[htbp]
    \vspace{-3cm}
    \centerline{\psfig{figure=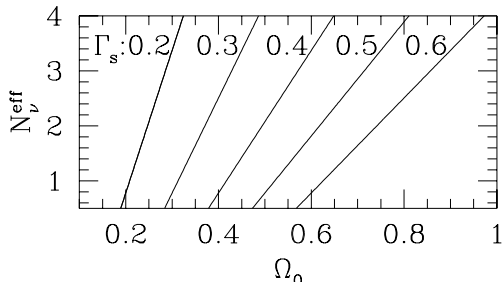,width=7cm}}

      \caption{
       Contours of $\Gamma_{\rm s} = 0.2$ (bold), $0.3, 0.4, 0.5$ and $0.6$ on 
       the ($\Omega_0, N_\nu^{\rm eff}$) plane for $h=0.7$.
      }
      \label{fig:cont_gam}
\end{figure}


Finally we discuss the CMB constraint on $T_R$. With the present angular
resolutions and sensitivities of COBE observation~\cite{COBE} or current
balloon and ground base experiments, it is impossible to set a
constraint on $N_{\nu}^{\rm eff}$. 
 However it is expected that future satellite
experiments such as MAP~\cite{MAP} and PLANCK~\cite{PLANCK} will give us
a useful information about $N_{\nu}^{\rm eff}$. From Lopez et al.'s
analysis~\cite{Lopez}, MAP and PLANCK have sensitivities that $\delta
N_{\nu}^{\rm eff} \gtrsim 0.1$ (MAP) and $0.03$ (PLANCK) 
including polarization
data, even if all cosmological parameters are determined simultaneously
(see also Fig.~5). From such future observations of anisotropies of CMB,
it is expected that we can precisely determine T$_R$.
\begin{figure}[htbp]
    \vspace{0cm}
    \centerline{\psfig{figure=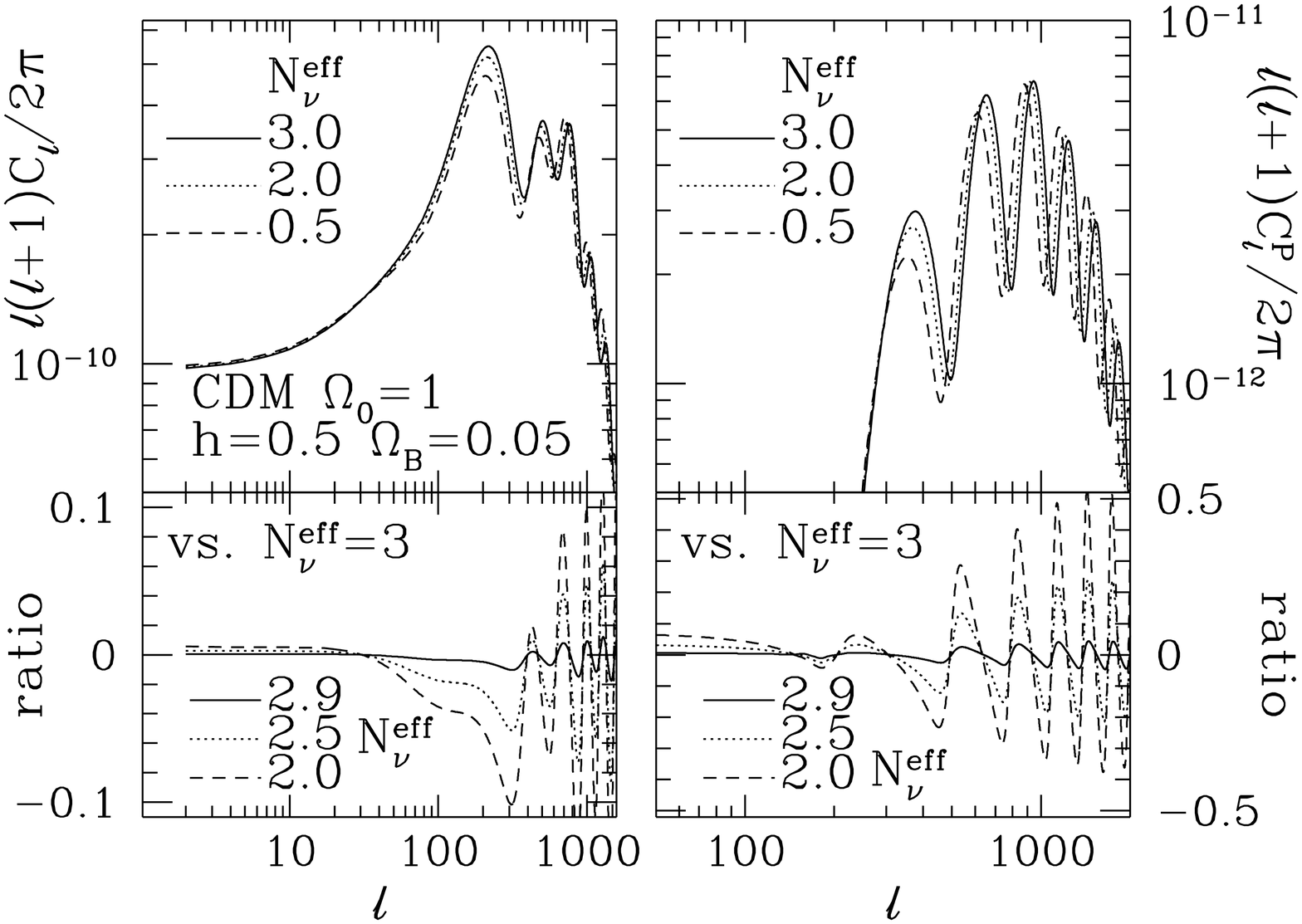,width=8.5cm}}

      \caption{
        Power spectra of CMB anisotropies (left top panel) and
       polarization (right top panel) of models with 
       $N_{\rm eff}^{\rm eff}=3, 2$
       and $0.5$.  Bottom two panels show $(C_\ell(N_{\rm eff})-
       C_\ell(3))/C_\ell(3)$ with $N_{\rm eff}^{\rm eff}=2.9, 2.5$
       and $2$ for CMB anisotropies (left bottom) and polarization (right 
       bottom).
       }
      \label{fig:cl_gam}
\end{figure}


\end{document}